\documentclass[acmtog,nonacm]{acmart}
\usepackage{subcaption}

\AtBeginDocument{%
  }

\setcopyright{none}

\usepackage[nameinlink,capitalise]{cleveref}
\crefname{line}{line}{lines}
\crefname{figure}{Fig.}{Figs.}
\Crefname{figure}{Fig.}{Figs.}
\crefname{equation}{Eq.}{Eqs.}
\Crefname{equation}{Eq.}{Eqs.}
\crefname{section}{Sec.}{Secs.}
\Crefname{section}{Sec.}{Secs.}
\crefname{definition}{Def.}{Defs.}
\Crefname{definition}{Def.}{Defs.}
\crefname{algorithm}{Alg.}{Algs.}
\Crefname{algorithm}{Alg.}{Algs.}
\crefname{assumption}{Asm.}{Asms.}
\Crefname{assumption}{Asm.}{Asms.}
\crefname{theoren}{Thm.}{Thms.}
\Crefname{theorem}{Thm.}{Thms.}
\crefname{subassumption}{Asm.}{Asms.}
\Crefname{subassumption}{Asm.}{Asms.}
\crefname{table}{Tab.}{Tabs.}
\Crefname{table}{Tab.}{Tabs.}

\begin{document}

\title{Evaluating Machine Learning Approaches for ASCII Art Generation}

\author{Sai Coumar}
\email{sai.c.coumar1@gmail.com}

\author{Zachary Kingston}
\affiliation{%
  \institution{Purdue University}
  \city{West Lafayette, IN}
  \country{USA}}
\email{zkingston@purdue.edu}

\begin{teaserfigure}
  \includegraphics[width=\textwidth]{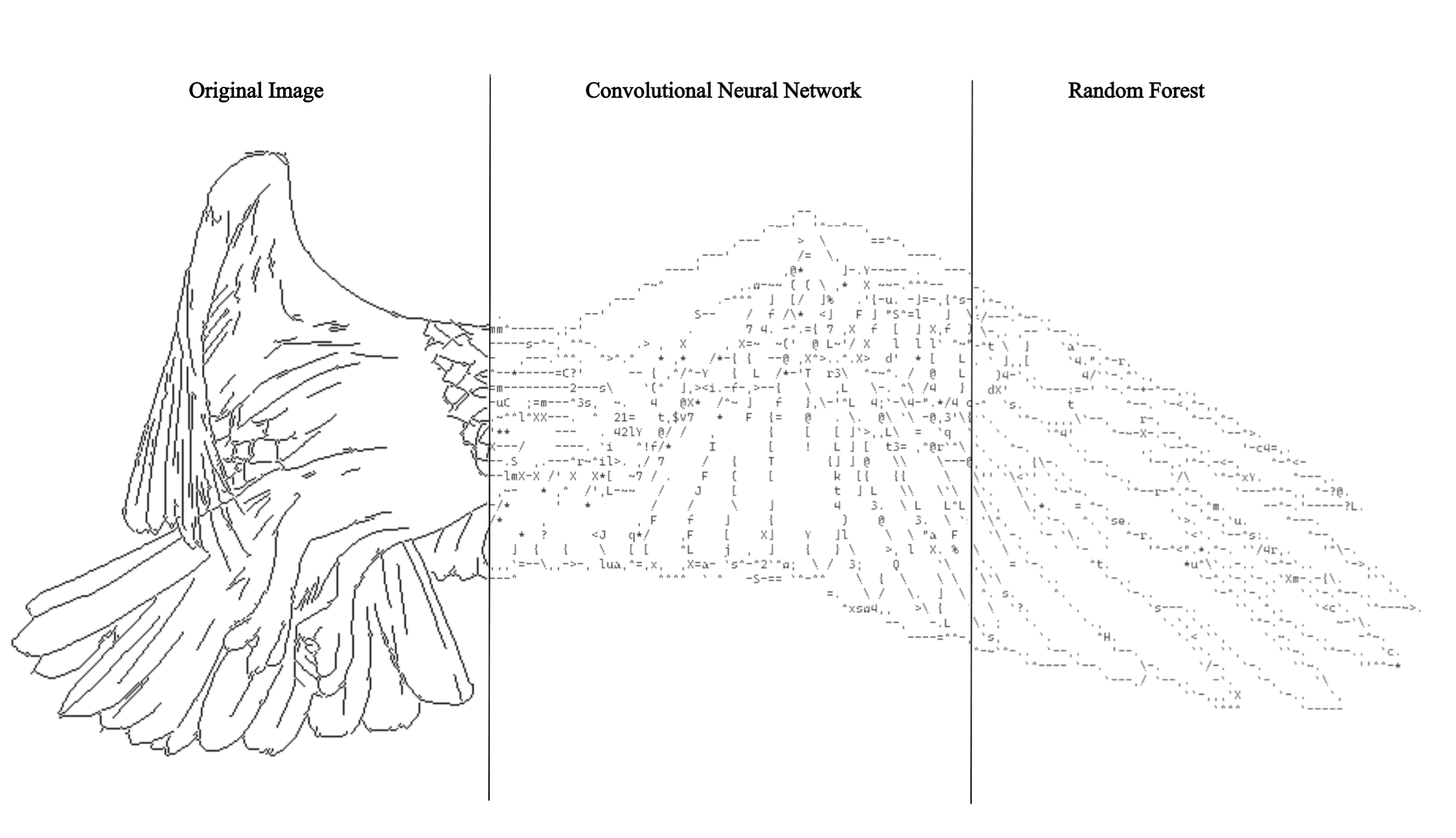}
  \caption{An image of an eagle converted into structure-based ASCII art using the CNN and Random Forest classifiers}
  \Description{An image of an eagle converted into structure-based ASCII art using the CNN and Random Forest classifiers}
  \label{fig:teaser}
\end{teaserfigure}

\renewcommand{\shortauthors}{Coumar et al.}

\begin{abstract}
Generating structured ASCII art using computational techniques demands a careful interplay between aesthetic representation and computational precision, requiring models that can effectively translate visual information into symbolic text characters. Although Convolutional Neural Networks (CNNs) have shown promise in this domain, the comparative performance of deep learning architectures and classical machine learning methods remains unexplored. This paper explores the application of contemporary ML and DL methods to generate structured ASCII art, focusing on three key criteria: fidelity, character classification accuracy, and output quality. We investigate deep learning architectures, including Multilayer Perceptrons (MLPs), ResNet, and MobileNetV2, alongside classical approaches such as Random Forests, Support Vector Machines (SVMs) and k-Nearest Neighbors (k-NN), trained on an augmented synthetic dataset of ASCII characters. Our results show that complex neural network architectures often fall short in producing high-quality ASCII art, whereas classical machine learning classifiers, despite their simplicity, achieve performance similar to CNNs. Our findings highlight the strength of classical methods in bridging model simplicity with output quality, offering new insights into ASCII art synthesis and machine learning on image data with low dimensionality.
\end{abstract}

\keywords{ASCII Art, Machine Learning}

\maketitle

\section{Introduction}
ASCII art is a type of digital art that uses text characters to create visual representations. Using the American Standard Code for Information Interchange (ASCII) character set~\cite{wiki:ASCII_art}, it produces visually interpretable media entirely in text form, making it suitable for display in any text-based environment. Renowned for its minimalist aesthetic, ASCII art transforms simple text into intricate designs that often carry a distinct charm and artistic appeal~\cite{dataartmuseum}. Although initially beginning as hand-crafted art forms, the intrinsic computational nature and near-universal support of the medium have led to the automation of synthetically generating ASCII art from natural photos. In addition to its cultural relevance, as Natural Language Processing and Large Language Model technologies continue to advance, the interpretation and generation of text-based visual art are becoming increasingly important challenges for automated tools to address~\cite{jiang2024artpromptasciiartbasedjailbreak}.

The process of converting an image to ASCII involves transforming the image into a text-based representation, where ASCII characters are chosen based on their visual resemblance to the intensity and brightness of the image's pixels. ASCII art can be categorized into two distinct types: tone-based and structure-based~\cite{10.1145/1833349.1778789}. Tone-based ASCII art involves replacing (pixels or segmented groups of pixels) with characters that match the intensity of the pixels, creating a gradient effect. In contrast, structure-based ASCII art arranges text characters to form contoured line structures, emphasizing the shape and outline of the image. This added emphasis on preserving and representing structural details makes structure-based ASCII art significantly more challenging, as it requires understanding of the image's geometry and spatial relationships.

Structured ASCII art synthesis was introduced by~\citet{10.1145/1833349.1778789} with the Alignment-Insensitive Shape Similarity (AISS) metric for ASCII character matching. More recently, learning approaches such as convolutional neural networks (e.g.,~\cite{Akiyama2017ASCIIAS}) have also been used. Despite proving that neural approaches are viable for ASCII art synthesis, there has been no comparison of the performance of different approaches towards ASCII art synthesis. 
Furthermore, the rapid advancement of hardware platforms~\cite{nvidia} and software support~\cite{pytorch, sckitlearn} facilitates access to a wide range of machine learning techniques for ASCII art synthesis, allowing flexible experimentation with complex machine learning models. 

In this paper, we evaluate various classical machine learning (ML) and deep learning (DL) methods to determine the most effective ML approaches for line structure replacement and further develop the role of machine learning in ASCII art generation. We compare k-Nearest Neighbors (k-NN)~\cite{knn}, Support Vector Machines (SVM)~\cite{svm}, and Random Forest Classifiers~\cite{randomforests}, both with and without Histogram of Gradient feature extraction~\cite{hog}, against Convolutional Neural Networks (CNN)~\cite{cnn}, ResNet~\cite{resnet}, and MobileNetV2~\cite{mobilenetv2}. Our results show that classical methods, particularly random forests, deliver competitive quality in ASCII art generation with significantly reduced computational overhead. Furthermore, we provide an open-source implementation of our conversion tool to support adoption and reproducibility\footnote{\url{https://github.com/saiccoumar/deep_ascii_converter}.}.

We set the criteria for success as being able to take an image and convert it into structure-based ASCII art comprised of text characters through Machine Learning techniques. To focus on this goal, we exclude methods such as diffusion~\cite{diffusion} or GANs~\cite{gan} that return an image that looks like it is made of ASCII text, rather than actually being ASCII text, as well as techniques that either create tone-based ASCII art.

\section{Related Work}

Examples of ASCII art can be found throughout the internet: the ASCII Art Archive~\cite{asciiart_archive} retains an extensive collection of human-made ASCII art, as well as its own tone-based ASCII converter. Additionally, the concept of synthesizing ASCII art has existed for some time; novel techniques for the conversion of images to structure-based ASCII art, such as AISS, use a similarity score between a group of pictures and a matching glyph, or ASCII character, and match the ones with the highest similarity score. AISS is particularly unique among similarity metrics in picking the same result regardless of rotation and translation. Glyph-matching methods have also been successfully employed in conjunction with other techniques to improve matching performance, such as feature extraction techniques such as the Histogram of Gradients (HoG) and Normalized Cross-Correlation (NCC)~\cite{Miyake2011AnIS}.

The first Convolutional Neural Network (CNN) approach marked an advance in ASCII art synthesis, achieving an estimated character classification accuracy of 89\%~\cite{Akiyama2017ASCIIAS} with well-defined structures. Accuracy of the model is evaluated by the percentage of tiles $n \times n$ correctly classified as ASCII characters in a labeled test set. Output images are evaluated both qualitatively, through observation of representative examples, and quantitatively, using the Structural Similarity Index Measure (SSIM)~\cite{SSIM} and Image2Vec Similarity (i2v)~\cite{i2v} scores. SSIM evaluates structural similarity, while i2v measures semantic similarity.

While CNNs have proven effective (e.g.,~\cite{8691245}), there is a lack of research into different architectures or the viability of simpler algorithms. An autoencoder-based approach~\cite{10.1145/3591569.3591587} was used to encode image segments into a latent space as a preprocessing step before character classification with k-NN. While this works well for tone-based art, it struggles with structure-based ASCII art and leads to shading, which does not adequately meet the demands of structure-based ASCII art. More notably, the k-NN classification provides a solid backbone for conversion, indicating potential for alternative ML approaches.  

\begin{figure*}[t]  %
    \centering
    \begin{subfigure}[t]{0.3\textwidth}
        \centering
        \includegraphics[width=\textwidth]{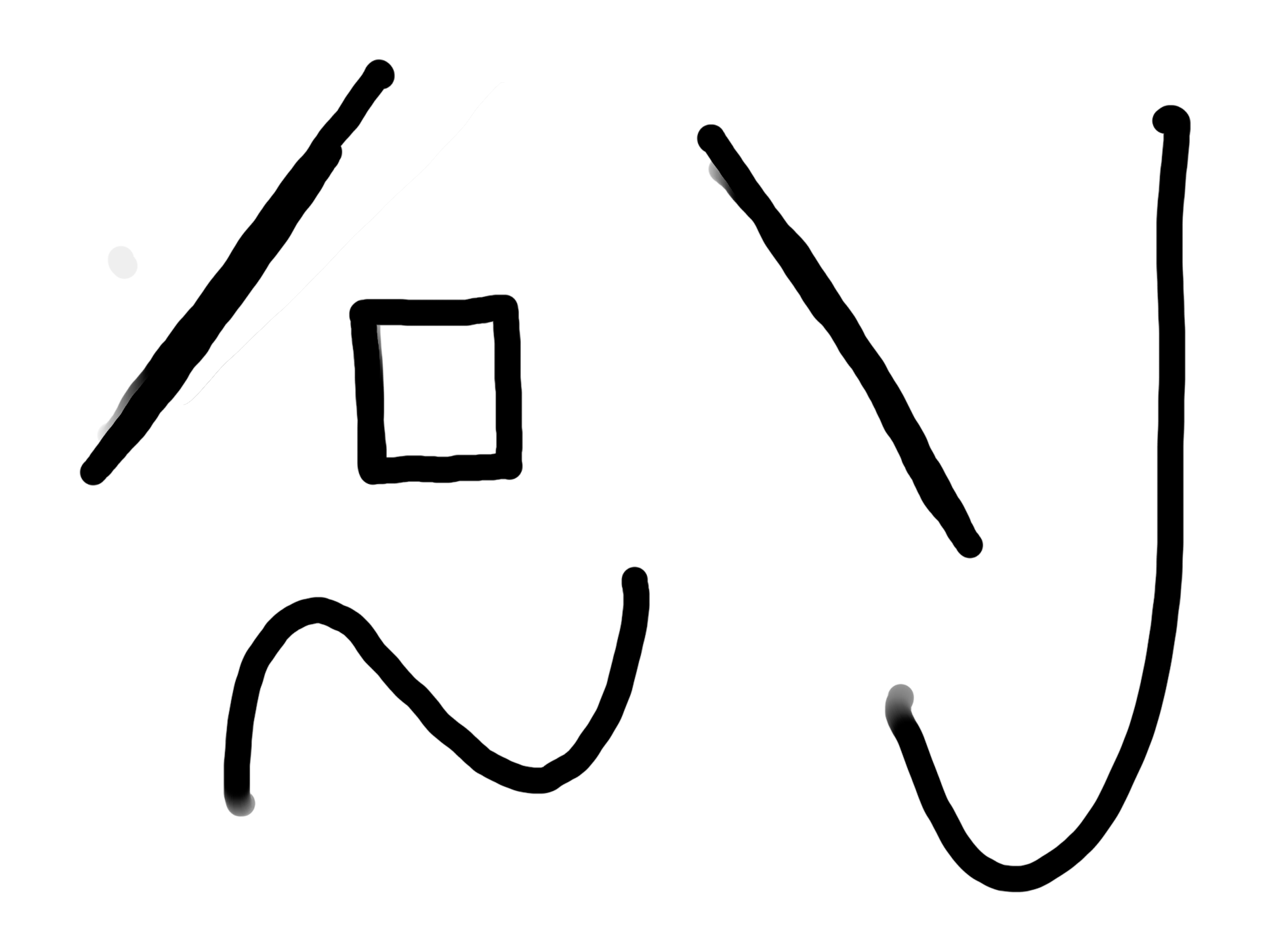}
        \caption{Original Image}
        \label{fig:original}
    \end{subfigure}
    \hfill
    \begin{subfigure}[t]{0.3\textwidth}
        \centering
        \includegraphics[width=\textwidth]{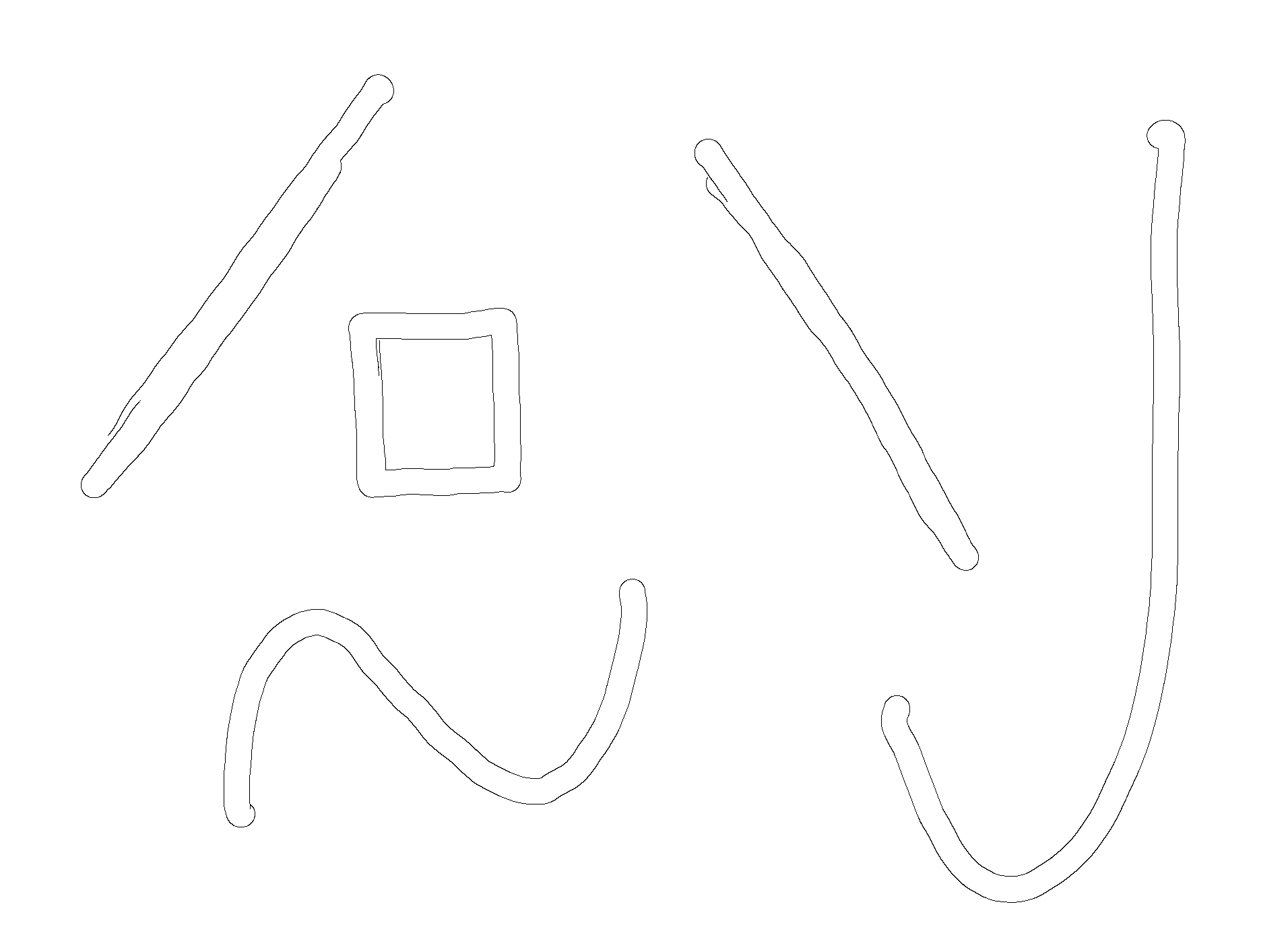}
        \caption{Extracted Structures}
        \label{fig:structures}
    \end{subfigure}
    \hfill
    \begin{subfigure}[t]{0.3\textwidth}
        \centering
        \includegraphics[width=\textwidth]{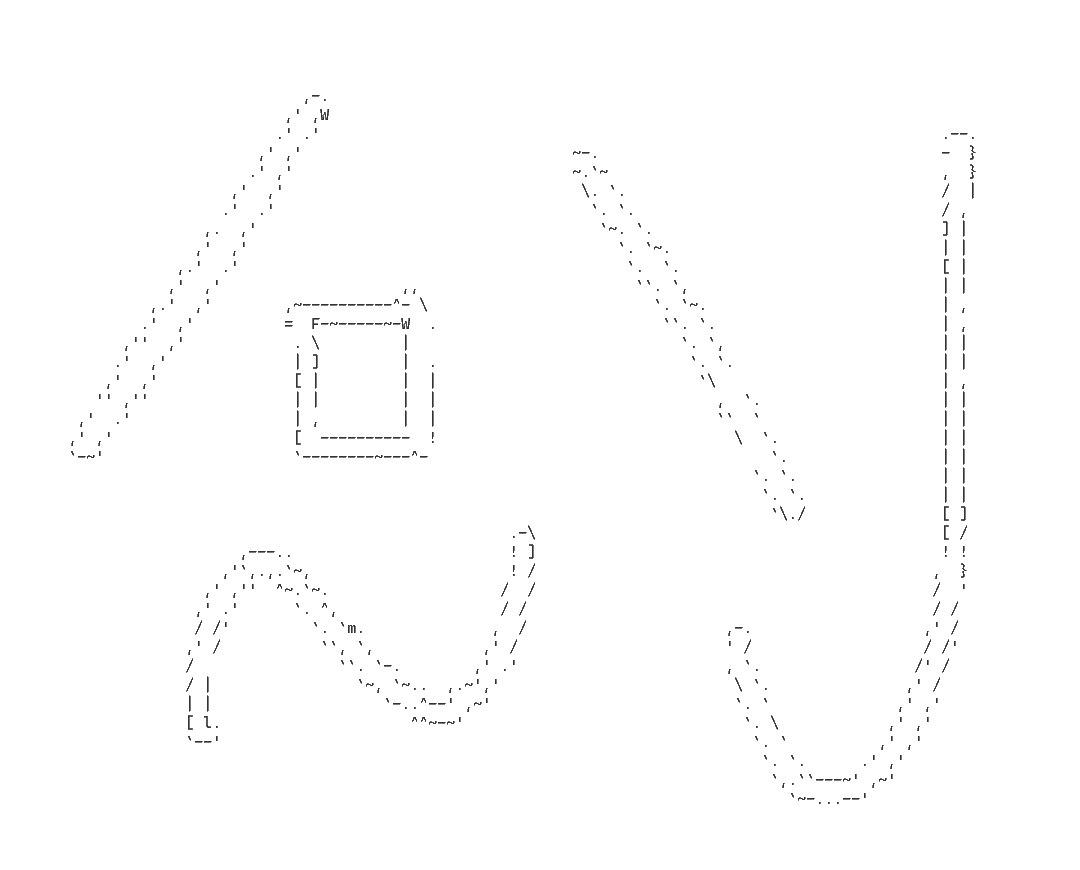}
        \caption{Final ASCII Conversion}
        \label{fig:ascii}
    \end{subfigure}
    \caption{Progression of ASCII Conversion: (a) Original image, (b) Intermediate structural extraction, (c) Final ASCII result (using k-NN).}
    \label{fig:ascii_conversion}
    \Description{Progression of ASCII Conversion: (a) Original image, (b) Intermediate structural extraction, (c) Final ASCII result (using k-NN).}
\end{figure*}

Techniques for image-to-ASCII art conversion generally involve two steps: extracting relevant line structures from the natural photograph and replacing these line structures with appropriate ASCII characters. As shown in ~\cref{fig:ascii_conversion}, this process includes segmenting the image into $n \times n$ tiles and matching the contents of each tile to an ASCII character.

\section{Methods} %
\subsection{Image Preprocessing} %

Line extraction is essential for converting natural photographs into ASCII characters. This can be done using methods like the Canny edge detector~\cite{canny} or non-CRF modulation~\cite{7491376}  or by using pre-extracted line structures. For efficiency, we use pre-extracted structures when comparing machine learning and deep learning methods, as this minimizes visual noise and accounts for potential imperfections in the extraction process.

Image resizing is done to rescale the image before ASCII conversion. Since text characters are not perfect squares, the height of an image must be reduced by a factor of $2$ so that the ASCII output does not appear stretched vertically. Rescaling also adds the functionality to output larger or smaller outputs based on a factor passed in as a parameter, and tune the output to improve structural replacement. Additionally, grayscaling is applied to the processed image by calculating a weighted sum of the color channels' luminance, as color information is not relevant to contour detection for conversion.

\begin{figure*}[ht]
    \centering
    \begin{tabular}{cc}
        \subfloat[k-NN using $10 \times  10$ tiles]{\includegraphics[width=0.45\linewidth]{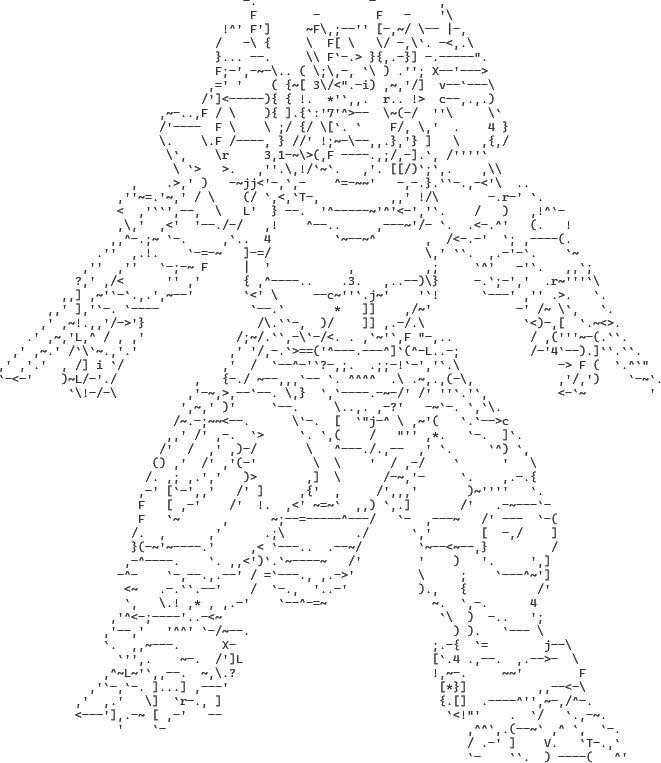}} &
        \subfloat[k-NN using $64 \times  64$ tiles]{\includegraphics[width=0.45\linewidth]{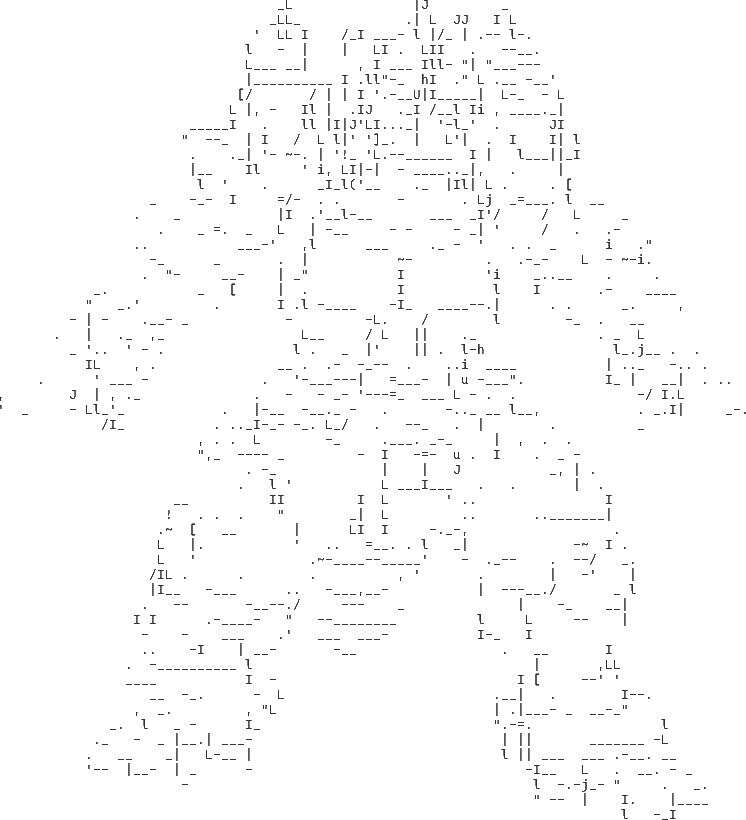}} \\
    \end{tabular}
    \caption{Effects of Tile Size on Image Quality}
    \label{fig:tile_size}
    \Description{Effects of Tile Size on Image Quality}
\end{figure*}

Following conversion to grayscale, the image is divided into ``tiles'', each of which is converted to an ASCII character. Although prior research~\cite{Akiyama2017ASCIIAS,10.1145/3591569.3591587} uses $64 \times  64$ tiles, we opted to use $10 \times  10$ to reduce the input dimensionality and make model training and inference more reasonable. This modification led to much lower inference times and marginally better visual outputs in small images with dense details. \Cref{fig:tile_size} displays an example where structures are much more defined using $10 \times  10$ tiles, specifically around areas such as the toes or knees. Additionally, the inference time for the output using $10 \times 10$ tiles was only 1.5055 seconds, while the output for $64 \times  64$ tiles had an inference time of 44.9350 seconds. Decreasing tile size makes experiments much more feasible, especially when using deep neural network architectures such as ResNet, which quickly run out of virtual memory during inference. 

\subsection{Tile Conversion} %
For a baseline tile conversion method, tile matching was performed using AISS~\cite{10.1145/1833349.1778789}. AISS computes log-polar histograms for tiles and minimizes the difference between them and log-polar histograms for ASCII characters to match a tile with the most similar ASCII character. This is quite effective despite being a simple glyph matching method, because log-polar histograms are inherently insensitive to small shape perturbations, leading to their misalignment-tolerance nature.

\subsubsection{Machine Learning Methods}
Classical machine learning models were trained on a random subset of the dataset with 2500 data entries, as increasing the size of the dataset past this point would increase the inference time with diminishing returns on the already high classification accuracy; deep learning methods were trained using the full dataset of 50,000 entries as they benefit from more training data. Each record contains a feature vector of the image and has a labeled number representing the decimal character code.

We evaluated three common classical machine learning models for supervised classification: SVM, Random Forest, and k-NN. The SVM was trained using a linear kernel, the Random Forest classifier with 100 estimators, and the k-NN classifier with 5 neighbors, all using the default scikit-learn~\cite{sckitlearn} parameters. Supplementary models were retrained using Histogram of Gradient (HoG) features from the original data. HoG methods were applied in~\cite{Miyake2011AnIS} with glyph matching using similarity metrics, and the benefits gained from using a derived feature extractor were evaluated for applications in ML models.

For deep learning models, various architectures were tested: the original CNN architecture outlined in~\citet{Akiyama2017ASCIIAS}, the Resnet18 architecture~\cite{resnet}, and the MobileNetV2 architecture~\cite{mobilenetv2}. Additionally, a standard Multilayer Perceptron (MLP)~\cite{mlp} architecture was tested to reference. All deep learning models were trained to convergence with a batch size of 256 and learning rate of 1e-3 for 10 epochs using Cross Entropy Loss with the Adam optimizer.

\section{Materials}
\subsection{Dataset Synthesis} %
Datasets for training models in character classification are difficult to obtain. Previous research~\cite{Akiyama2017ASCIIAS} collected ASCII art, recreated the line structures, and segmented the images into tiles, which were then used to train the character classifier. Instead of recreating the data, we synthesized a dataset by generating image tiles of ASCII characters and associating each character with its corresponding image. This approach was also used for prior research collecting data for the autoencoder preprocessor~\cite{10.1145/3591569.3591587}.

The dataset was augmented by taking random samples and applying transformations in order to create more samples for a given character; transformations included were Gaussian blurring, positional shifts, and random noise. A CNN character classifier~\cite{Akiyama2017ASCIIAS} was successfully recreated using our synthetic dataset instead of manual collection and estimation, and as such we consider it a valid substitute when training with other machine learning techniques.

While methods exist to efficiently generate structured ASCII art with various character sets~\cite{fasttextplacement}, we narrowed our dataset to include only the original ASCII character set in order to more closely compare how different modeling techniques compare in structured ASCII art generation.

\subsection{Hardware Platforms}
Models were trained on a PC with a 13th Gen Intel(R) Core(TM) i7-13700K CPU, an NVIDIA GeForce RTX 4070Ti, and 32GB of system memory. Hardware acceleration was used for model training and prediction throughout. 

Deep learning techniques were implemented using PyTorch to exploit hardware acceleration, and classical machine learning techniques were implemented using the scikit-learn~\cite{sckitlearn} package.

\section{Results}
\begin{figure*}[ht]
    \centering
    \begin{tabular}{cc}
        \subfloat[k-NN]{\includegraphics[width=0.45\linewidth]{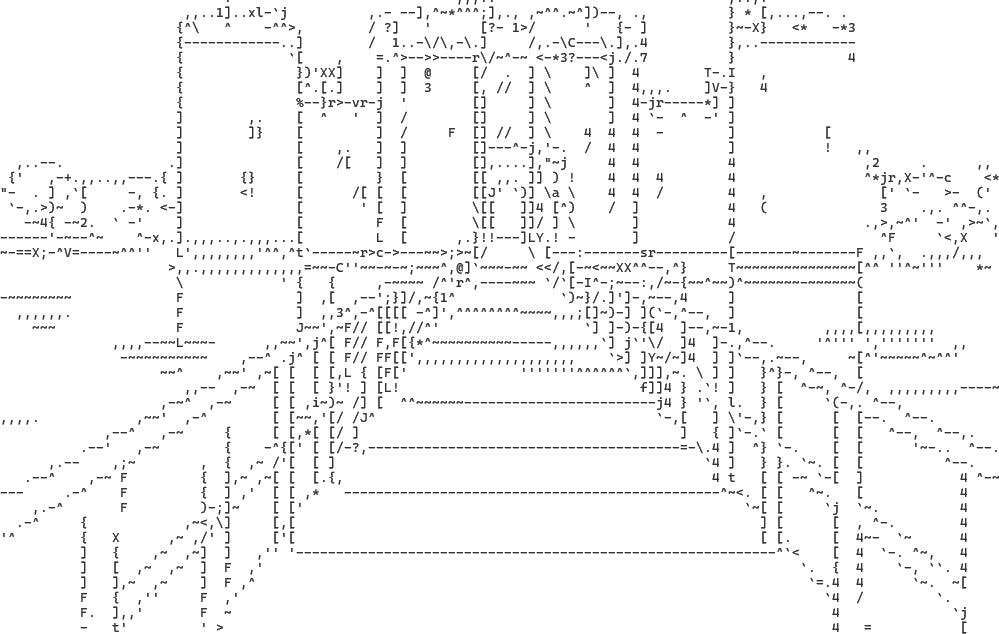}} &
        \subfloat[k-NN with Autoencoder Preprocessing~\cite{10.1145/3591569.3591587}]{\includegraphics[width=0.45\linewidth]{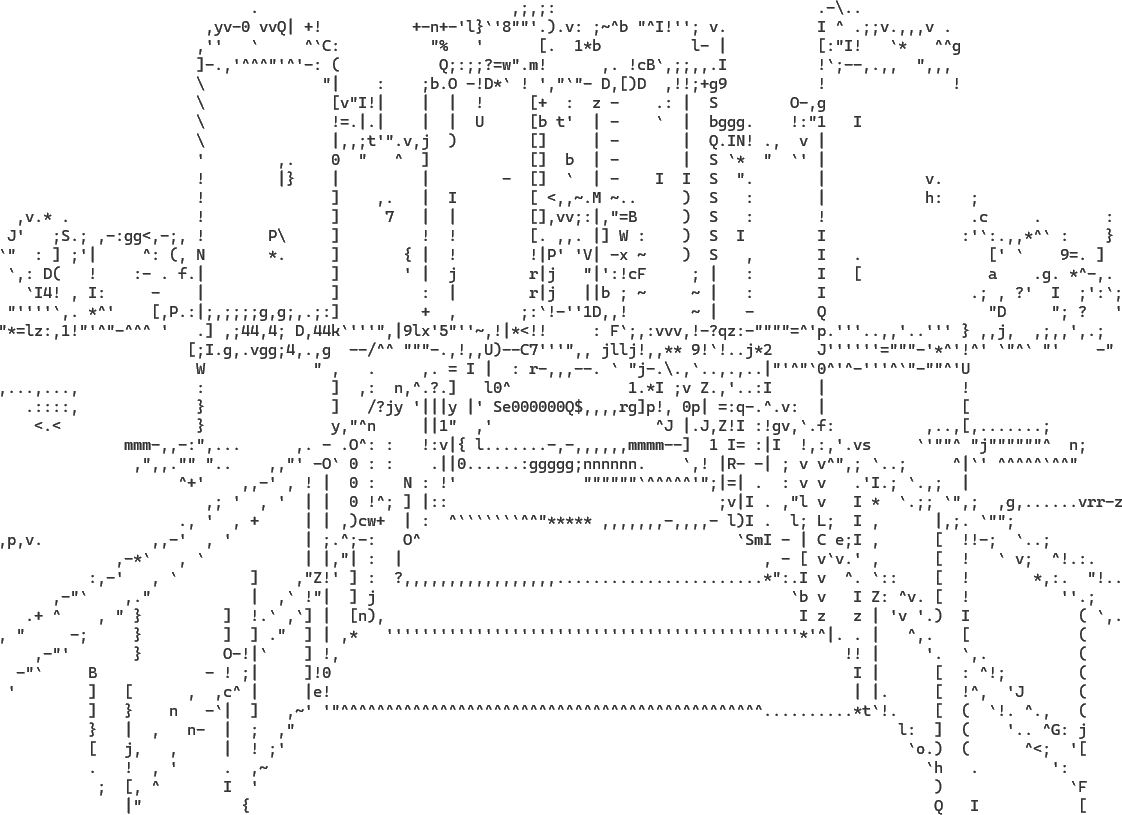}}\label{fig:2} \\
    \end{tabular}
    \caption{Autoencoder preprocessing decreases structural fidelity and creates overmatching}
    \label{fig:autoenc}
    \Description{Autoencoder preprocessing decreases structural fidelity and creates overmatching}
\end{figure*}

We first evaluate the viability of preprocessing the input data with an autoencoder prior to feeding the feature vector to the character classifiers. This approach uses an autoencoder to preprocess the input image by extracting latent features, effectively reducing irrelevant information. The latent features serve as the input for a k-NN classifier, which subsequently maps the features to ASCII characters. This combination aims to enhance the classification process by transforming raw input into a feature space that k-NN can use more effectively.

Unfortunately, the autoencoder in this pipeline is not beneficial when applied to structured ASCII art. As shown in ~\cref{fig:autoenc}, the outputs generated by k-NN with and without autoencoder preprocessing exhibit minimal qualitative benefits and degrade line structures, decreasing the i2v score from 0.66 to 0.6317. This suggests that the latent features extracted by the autoencoder do not contribute significantly to improving the structural accuracy or aesthetic quality of ASCII art. Instead, the results heavily rely on the effectiveness of the k-NN classifier in accurately associating input tiles with corresponding ASCII characters.

The inefficacy of the autoencoder can be attributed to an emphasis on preserving contour and structure in ASCII art, which may not benefit substantially from high-level feature extraction. Consequently, this underscores the importance of optimizing the classification process itself, rather than relying on complex preprocessing that does not improve the quality of the output.

\begin{table}
  \caption{Training and Test Accuracy for Character Classification Models}
  \label{tab:training_stats}
  \begin{tabular}{ccc}
    \toprule
    Model & Training Acc. (\%) & Test Acc. (\%)  \\
    \midrule
    k-NN & 96.9\% & 95.7\% \\
    k-NN w. HoG & 91.8\% & 85.1\% \\
    SVM & 94.8\% & 93.8\% \\
    SVM w. HoG & 96.9\% & 94.5\% \\
    Random Forest & 98.1\% & 91.4\% \\
    Random Forest w. HoG & 98.0\% & 95.0\% \\
    Neural Network & 35.9\% & 35.9\% \\
    CNN & 88.6\% & 96.0\% \\
    ResNet & \textbf{96.9}\% & \textbf{96.8}\% \\
    MobileNetV2 & 96.2\% & 96.3\% \\
    \bottomrule
  \end{tabular}
\end{table}

\begin{table}
  \caption{Metrics for Classical Machine Learning Models (with and without HoG)}
  \label{tab:classical_ml_metrics}
  \begin{tabular}{lcccc}
    \toprule
    Model & F1 Score & Recall \\
    \midrule
    k-NN  & \textbf{0.95} & \textbf{0.96} \\
    k-NN (with HoG) & 0.85 & 0.85 \\
    SVM & 0.93 & 0.94 \\
    SVM (with HoG) & 0.94 & 0.94 \\
    Random Forest & 0.91 & 0.91 \\
    Random Forest (with HoG) & 0.94 & 0.95 \\
    \bottomrule
  \end{tabular}
\end{table}
\begin{figure*}[ht]
    \centering
    \begin{tabular}{ccccc}
        \subfloat[Original Image]{\includegraphics[width=0.18\linewidth]{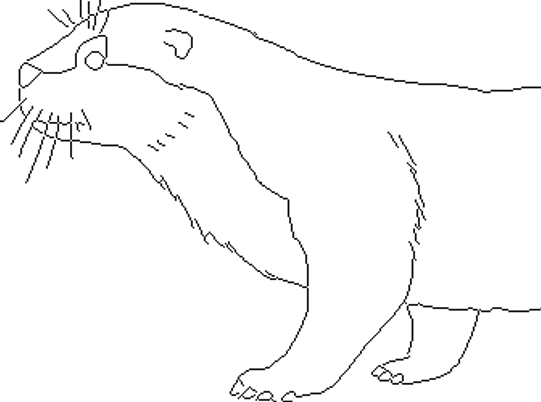}\label{fig:full_compare.1}} &
        \subfloat[AISS]{\includegraphics[width=0.18\linewidth]{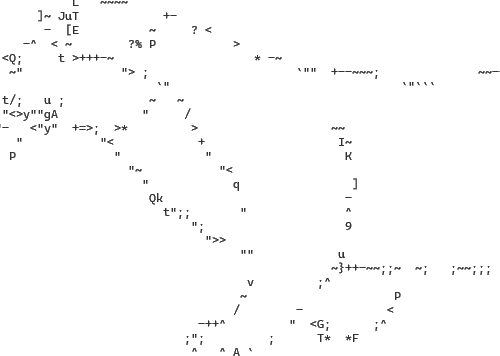}\label{fig:full_compare.2}} &
        \subfloat[Neural Network]{\includegraphics[width=0.18\linewidth]{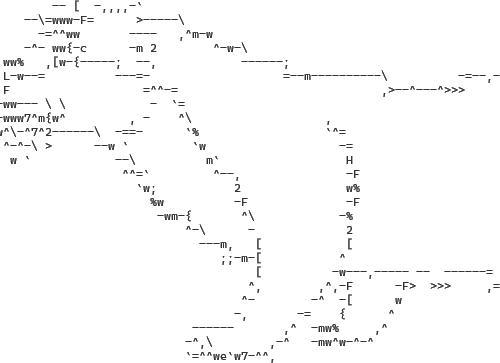}\label{fig:full_compare.3}} &
        \subfloat[k-NN]{\includegraphics[width=0.18\linewidth]{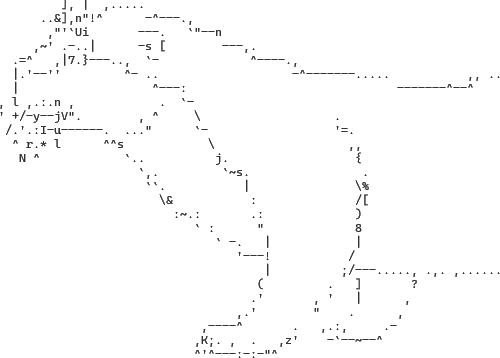}\label{fig:full_compare.4}} &
        \subfloat[SVM]{\includegraphics[width=0.18\linewidth]{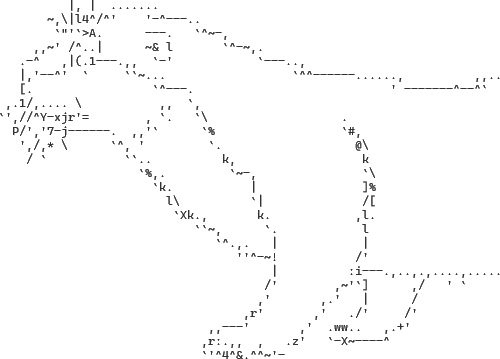}\label{fig:full_compare.5}} \\

        \subfloat[Random Forest]{\includegraphics[width=0.18\linewidth]{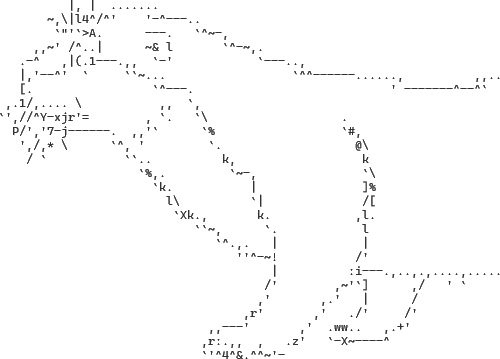}\label{fig:full_compare.6}} &
        \subfloat[CNN]{\includegraphics[width=0.18\linewidth]{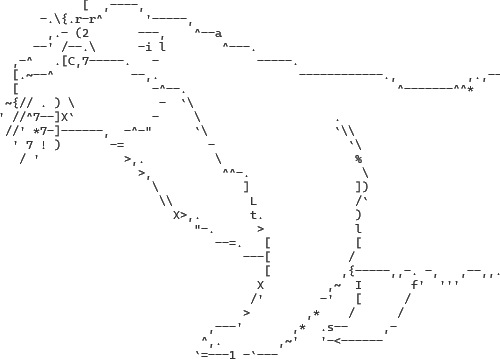}\label{fig:full_compare.7}} &
        \subfloat[ResNet]{\includegraphics[width=0.18\linewidth]{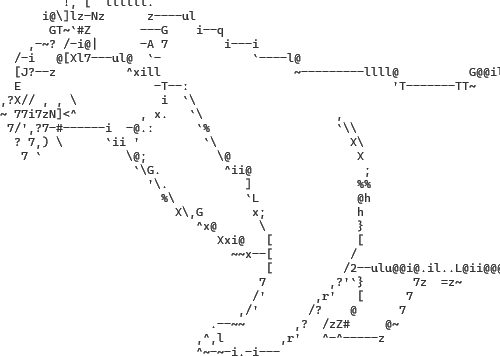}\label{fig:full_compare.8}} &
        \subfloat[MobileV2]{\includegraphics[width=0.18\linewidth]{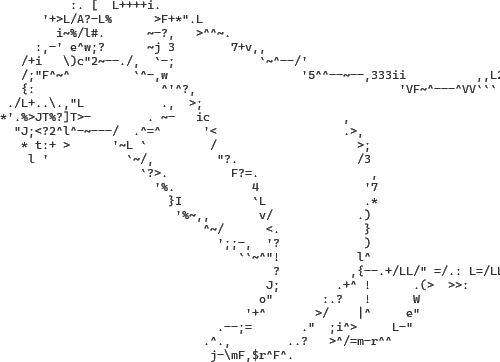}\label{fig:full_compare.9}} &
        \subfloat[AAConverter]{\includegraphics[width=0.18\linewidth]{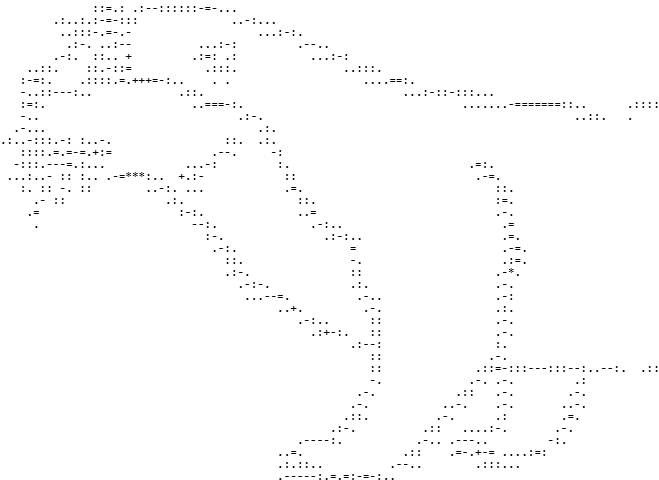}\label{fig:full_compare.10}} \\
    \end{tabular}
    \caption{Comparing various techniques for character classification shows that Random Forest matches CNNs in visual output quality}
    \Description{Comparing various techniques for character classification shows that Random Forest matches CNNs in visual output quality}
    \label{fig:full_compare}
\end{figure*}

Observing the accuracy of the models, all seem to perform extremely well. During training, all models achieved accuracy higher than $94\% $ except the simple CNN architecture, which had $88.6\%$ accuracy, and the basic neural network approach, which achieved only $35.9\%$. The model accuracy for the CNN method is consistent with the results of~\citet{Akiyama2017ASCIIAS}, and a qualitative visual comparison with our implementation shows that the final results are consistent, although our implementation uses the standard ASCII character set as opposed to the Japanese character set, ShiftJIS. On the test set, almost all models had above $90\%$ accuracy; SVM and Random Forest were able to achieve this with F1 and recall scores above $92\%$. The only exception was the classic neural network approach, which stood out with a low test accuracy of $35.9\%$.

Despite low model accuracy, the MLP approach still produced successful results. Although the results have some clutter in highly detailed areas, there is evidence of structure in more distinct areas. In the example images provided, the long straight lines are often replaced with the character ``-'', which develops more structure than in a novel non-ML technique like AISS. This indicates that model accuracy could potentially be a poor indicator for successful ASCII art synthesis.

Additionally, ResNet and MobileNetV2 had extremely high model test accuracies but suffered from ``overmatching''~\cite{fasttextplacement}, which is defined as when other characters that are not suitable may be matched. In ~\cref{fig:full_compare}, this is particularly noticeable in areas with dense visual information, such as the eyes and mouth.

\begin{table}[ht]
  \caption{SSIM, i2v Similarity, and Execution Times for Different Techniques}
  \label{tab:similarity}
  \centering
  \resizebox{\columnwidth}{!}{%
  \begin{tabular}{lccc}
    \toprule
    Technique & SSIM & i2v  & Conversion Time (ms) \\
    \midrule
    Original Image          & 1.0000 & 100.00 & - \\
    AISS                    & 0.6681 & 67.60  & 2931.37 \\
    AAConverter             & 0.6317 & 63.38  & - \\
    Neural Network          & 0.6342 & 71.15  & 267.68 \\
    k-NN                    & 0.6600 & 75.66  & 264.16 \\
    SVM                     & 0.6466 & 72.58  & 4630.18 \\
    Random Forest           & \textbf{0.6654} & 76.77 & \textbf{152.71} \\
    k-NN (with HoG)         & 0.6641 & 73.82  & 1291.95 \\
    SVM (with HoG)          & 0.6459 & 74.58  & 9468.78 \\
    Random Forest (with HoG)& 0.6549 & 76.36  & 1095.81 \\
    k-NN (with Autoencoder) & 0.6317 & 70.08  & 266.51 \\
    CNN                     & 0.6638 & \textbf{76.79} & 262.30 \\
    ResNet                  & 0.6364 & 72.98  & 289.94 \\
    MobileV2                & 0.6333 & 71.59  & 264.14 \\
    \bottomrule
  \end{tabular}%
  }
\end{table}

The SSIM and i2v results highlight the trade-off between structural accuracy (SSIM) and semantic fidelity (i2v). SSIM measures how closely the generated ASCII art matches the structure of the original image, while i2v evaluates semantic similarity by considering the visual content represented. The novel technique AISS, which does not use machine learning, achieved an i2v score of 67.60, indicating low semantic similarity, but achieves the highest SSIM with 0.6681, which indicates high structure retention. Despite the high structural accuracy, the low semantic similarity indicates that the model was unsuccessful at ASCII conversion while retaining the essence of the image. Holistically evaluating each model that meets both metrics can indicate the success of converting the original image to ASCII art. 

The CNN approach achieved balanced performance, with an SSIM of 0.6638 and an i2v of 76.79, indicating strong structural and semantic preservation, with an execution time of 262.30 milliseconds. Random Forest achieved a similar i2v score of 76.77, along with an SSIM score of 0.6654---slightly higher than the CNN's---suggesting that the Random Forest character classifier can at least match and even occasionally outperform CNNs, despite being a simpler model that requires only 152.71 milliseconds on average to classify the image into ASCII characters, 109.59 milliseconds less than the CNN.

In particular, the SSIM scores for ResNet and MobileNetV2 were among the lowest (0.6364 and 0.6333, respectively), emphasizing the structural inaccuracies caused by overmatching. Their corresponding i2v scores (72.98 and 71.59) further reinforce the inability of deeper models to balance semantic fidelity and structural accuracy. The phenomenon of overmatching could be attributed to the well-documented loss of detailed spatial information in higher layers of deep convolutional neural network architectures~\cite{styletransfer}. As each tile is only $10 \times 10$ or $64 \times 64$, the input dimensionality is extremely low, and the value of each pixel is higher than it would be in an image with larger dimensionality, which could make discerning between characters much more difficult for deep learning models that lose spatial information in deeper layers.
\begin{figure*}[ht]
    \centering
    \begin{tabular}{ccc}
        \subfloat[k-NN]{\includegraphics[width=0.3\linewidth]{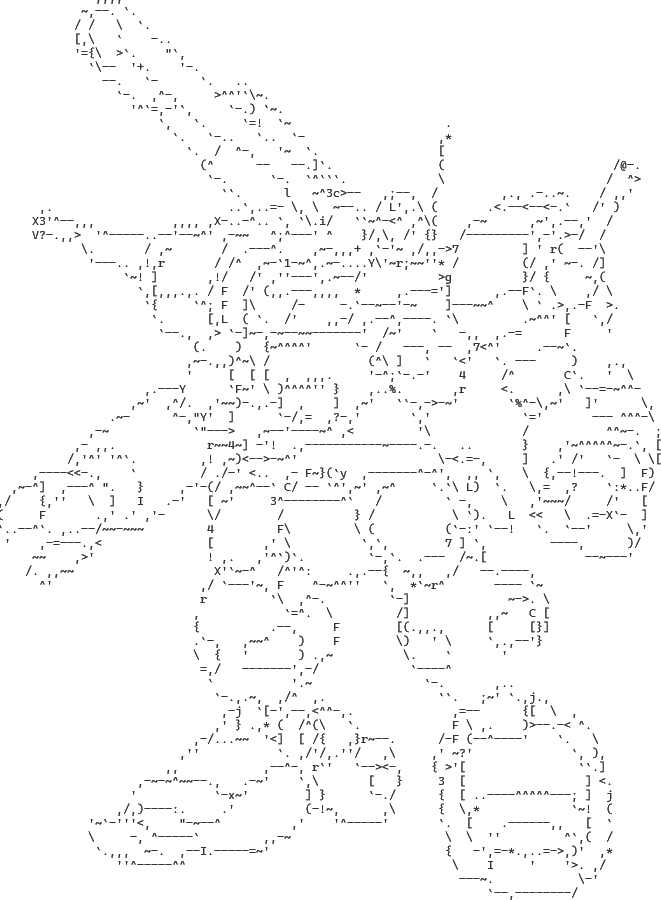}\label{fig:hog.4}} &
        \subfloat[SVM]{\includegraphics[width=0.3\linewidth]{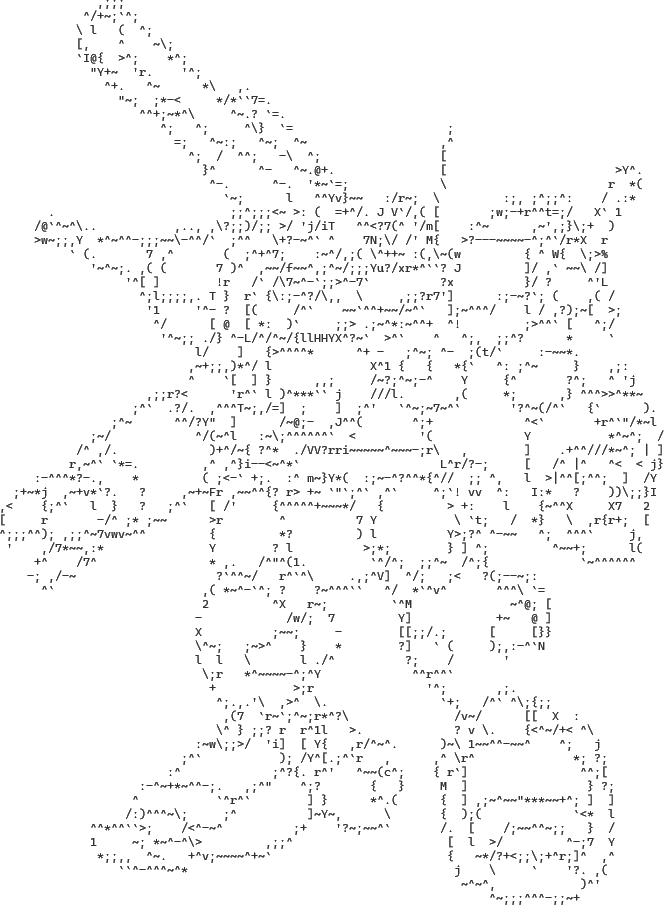}\label{fig:hog.5}} &
        \subfloat[Random Forest]{\includegraphics[width=0.3\linewidth]{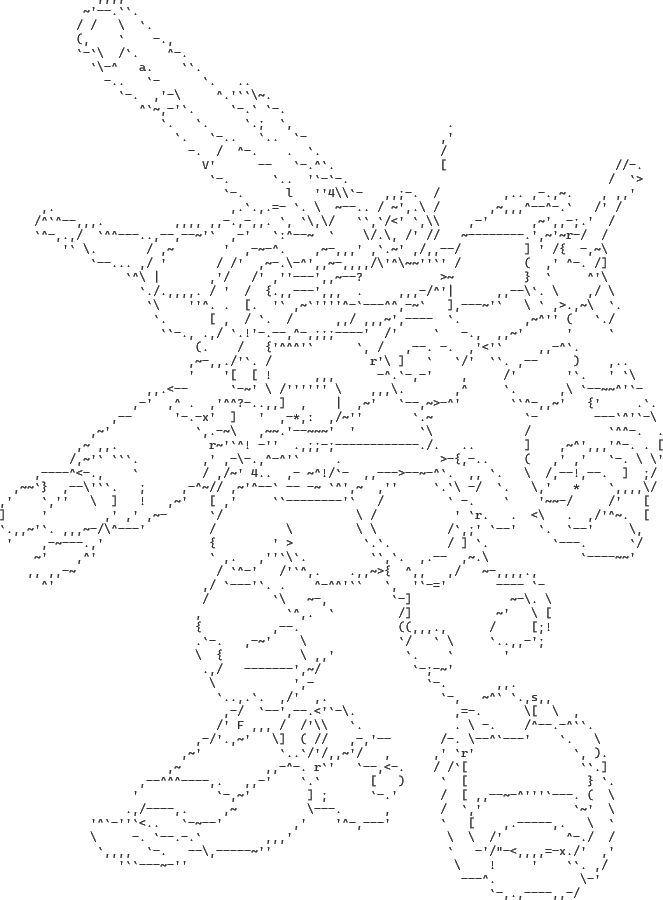}\label{fig:hog.6}} \\\\

        \subfloat[k-NN with HoG]{\includegraphics[width=0.3\linewidth]{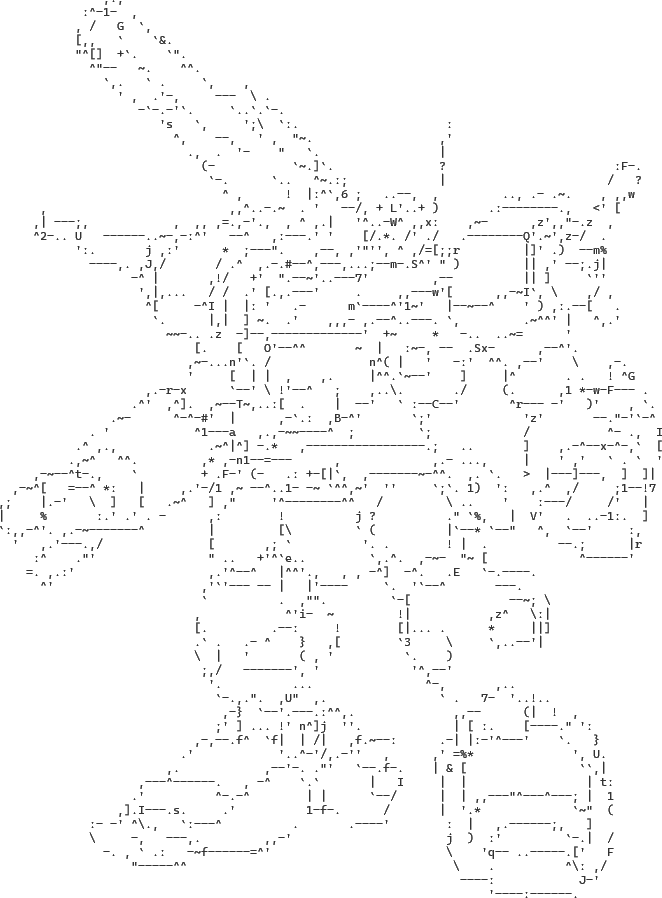}\label{fig:hog.1}} &
        \subfloat[SVM with HoG]{\includegraphics[width=0.3\linewidth]{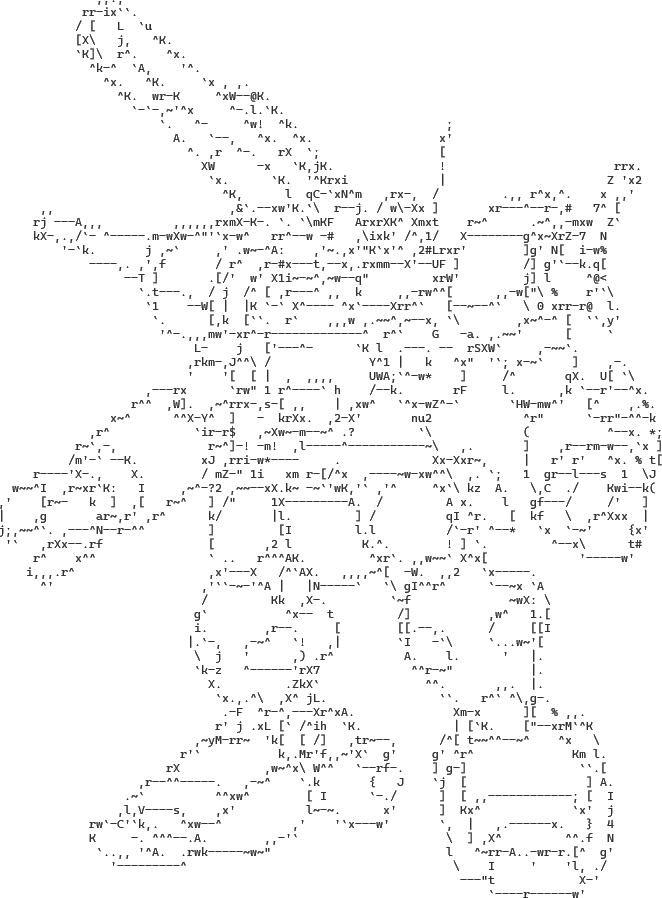}\label{fig:hog.2}} &
        \subfloat[Random Forest with HoG]{\includegraphics[width=0.3\linewidth]{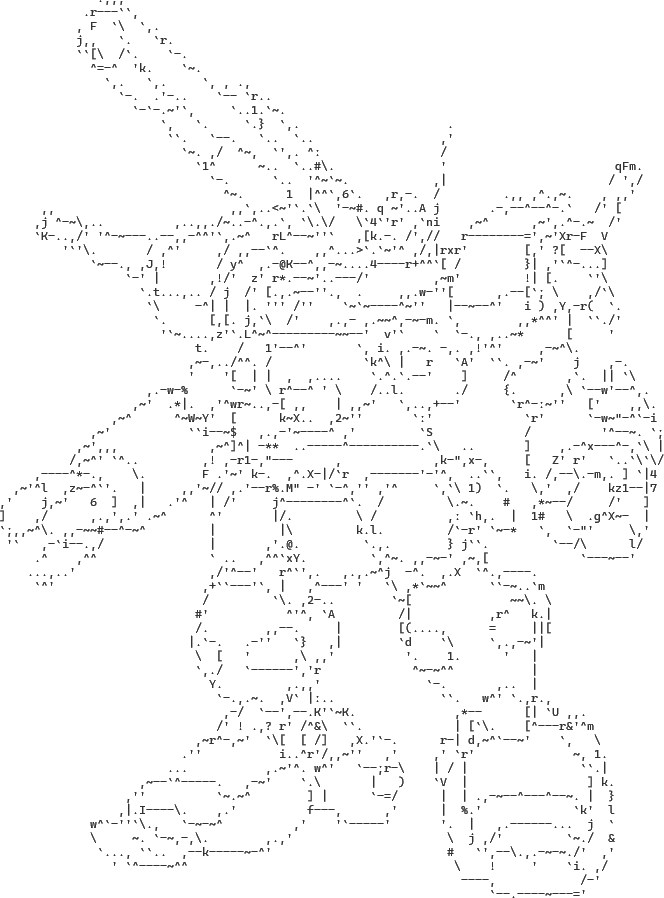}\label{fig:hog.3}} \\\\
        
    \end{tabular}
    \caption{Histogram of Gradients does not significantly affect output quality for classical ML models.}
    \Description{Histogram of Gradients does not significantly affect output quality for classical ML models.}
    \label{fig:hog}
\end{figure*}

Qualitative examination supports these quantitative results, as the output shows that classical ML algorithms such as k-NN and Random Forest with HoG match the quality found from CNNs despite being much simpler modeling techniques. When comparing the art generated from the Random Forest with that of the CNN, the output is extremely similar, and we can occasionally observe better structure development in areas with dense detail, such as the face in ~\cref{fig:full_compare}. SVM stands out as having uniquely subpar performance compared to k-NN and Random Forest due to suffering from overmatching similar to ResNet and MobileNetV2, accompanied by a slow execution time. Furthermore, ~\cref{fig:hog} shows that the modifications made using HoG features are minimal and do not provide a significant benefit to the definition of the structure. Additionally, the results in \cref{tab:similarity} show that utilizing HoG features marginally reduces SSIM and i2v similarity, reducing both structure and semantic fidelity. All methods have a better-defined structure than novel techniques like AISS or popular ASCII converts like the AAConverter.

Overall, the results demonstrate that models with moderate complexity, such as CNN and Random Forest, achieve balanced SSIM and i2v scores and are better suited for ASCII art synthesis, effectively capturing both structural and semantic qualities while avoiding the pitfalls of overmatching.

\section{Conclusion}

This study highlights the surprising efficacy of classical machine learning methods, particularly Random Forest, in structured ASCII art synthesis. Random Forest consistently matched the CNN approach in generating structurally accurate and aesthetically coherent ASCII art. This performance emphasizes the strength of simpler, interpretable methods in preserving essential details, even in a computationally constrained task like ASCII art generation.

In contrast, deep learning models exhibited notable limitations. The ``overmatching'' phenomenon, where models misclassify characters in dense or visually complex regions, highlighted the difficulty deep architectures face in managing low-dimensional input data and maintaining spatial precision. These findings suggest that, in domains with unique low-dimensionality like ASCII art synthesis, deeper networks may not always be the optimal choice, particularly when clarity and structural fidelity are prioritized.

This work challenges the assumption that deeper models always yield better results, advocating a more nuanced approach to model selection based on task-specific requirements. Future research could explore innovative model combinations that leverage the precision of Random Forest along with novel domain-specific optimization techniques, such as mismatch scores~\cite{10.1145/1833349.1778789} or expanded character sets~\cite{fasttextplacement}, paving the way for more effective ASCII art generation techniques. Furthermore, the open-source implementation and detailed investigation into machine learning techniques presented here provide a baseline for further exploration into ML for ASCII synthesis.

\bibliographystyle{ACM-Reference-Format}
\bibliography{references}
\begin{figure*}[htp]
    \centering
    \begin{tabular}{cc}
        \subfloat[Original Image]{\includegraphics[width=0.45\linewidth]{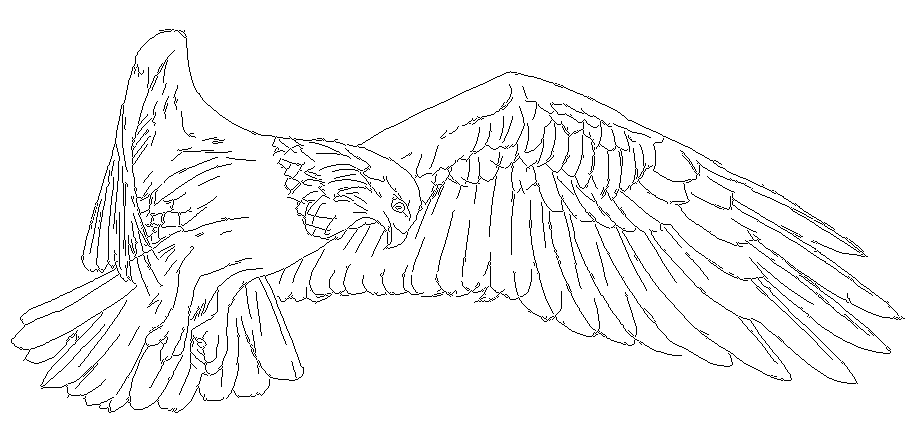}\label{fig:full_compare_bird.1}} &
        \subfloat[AISS]{\includegraphics[width=0.45\linewidth]{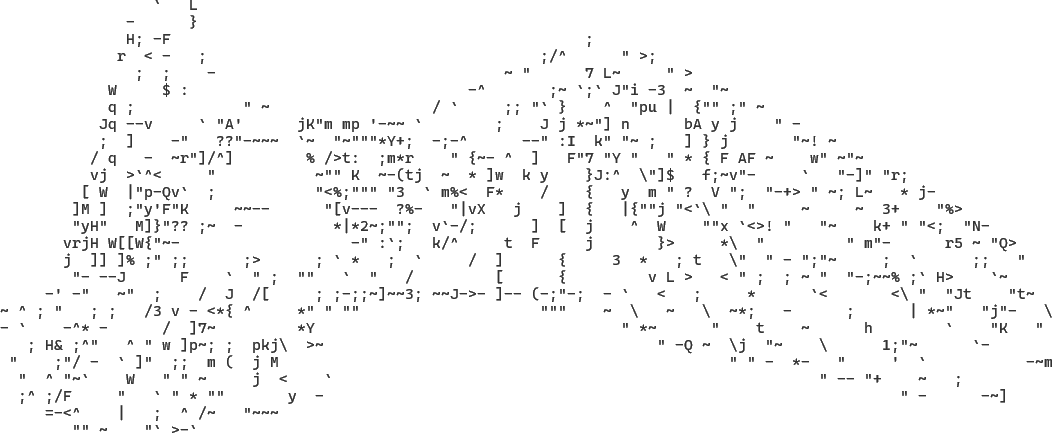}\label{fig:full_compare_bird.2}} \\
        \subfloat[Neural Network]{\includegraphics[width=0.45\linewidth]{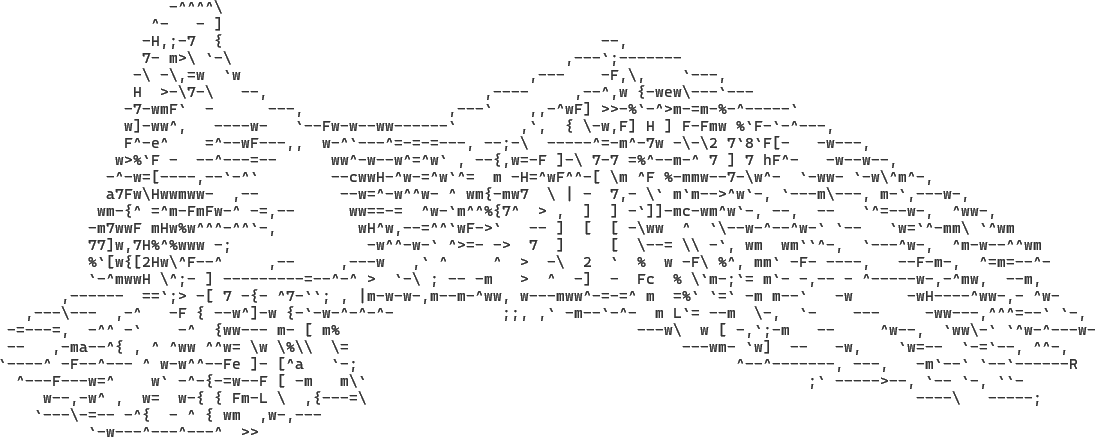}\label{fig:full_compare_bird.3}} &
        \subfloat[k-NN]{\includegraphics[width=0.45\linewidth]{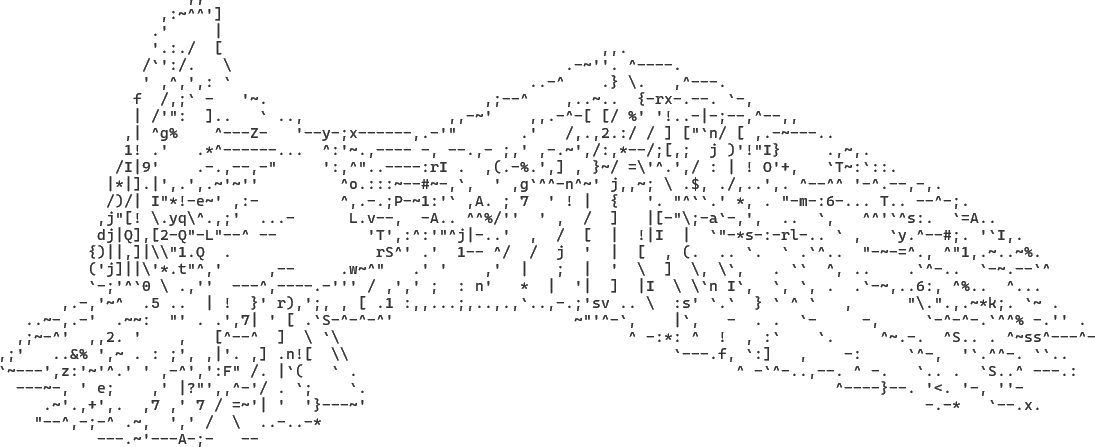}\label{fig:full_compare_bird.4}} \\
        \subfloat[SVM]{\includegraphics[width=0.45\linewidth]{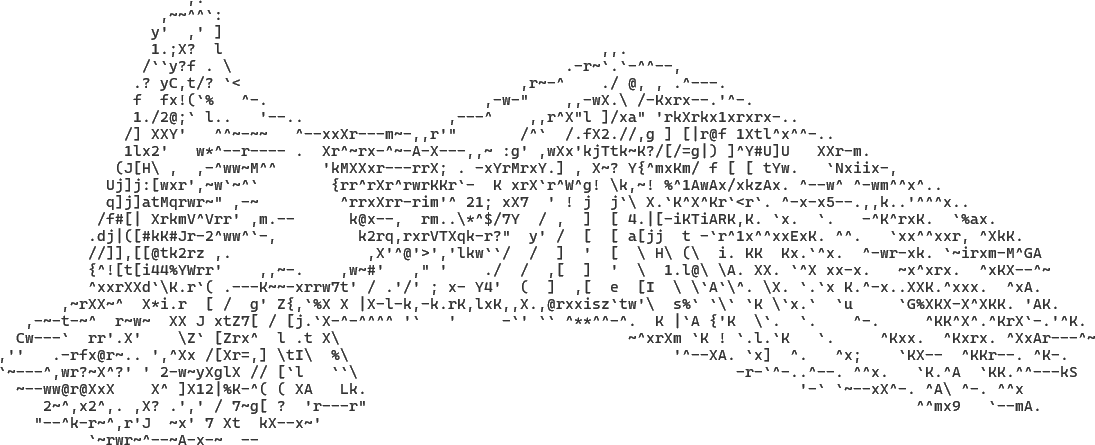}\label{fig:full_compare_bird.5}} &
        \subfloat[Random Forest]{\includegraphics[width=0.45\linewidth]{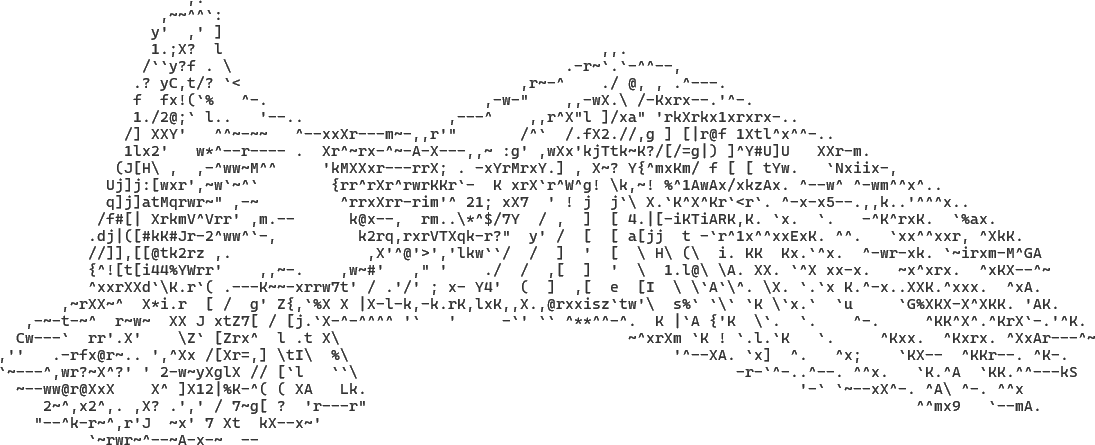}\label{fig:full_compare_bird.6}} \\
        \subfloat[CNN]{\includegraphics[width=0.45\linewidth]{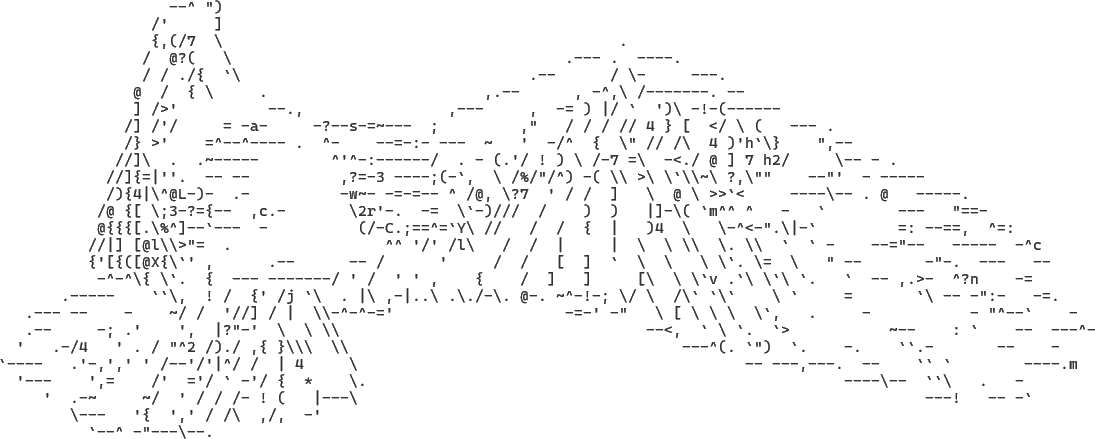}\label{fig:full_compare_bird.7}} &
        \subfloat[ResNet]{\includegraphics[width=0.45\linewidth]{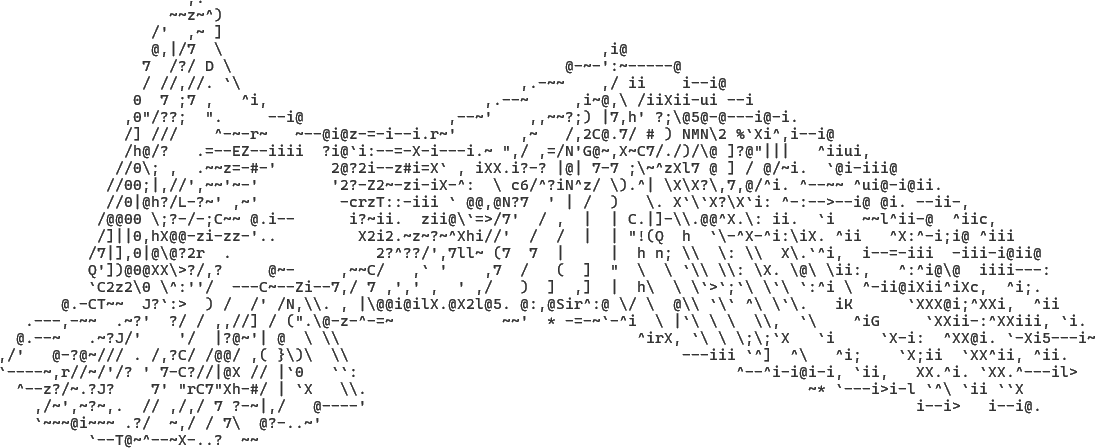}\label{fig:full_compare_bird.8}} \\
        \subfloat[MobileV2]{\includegraphics[width=0.45\linewidth]{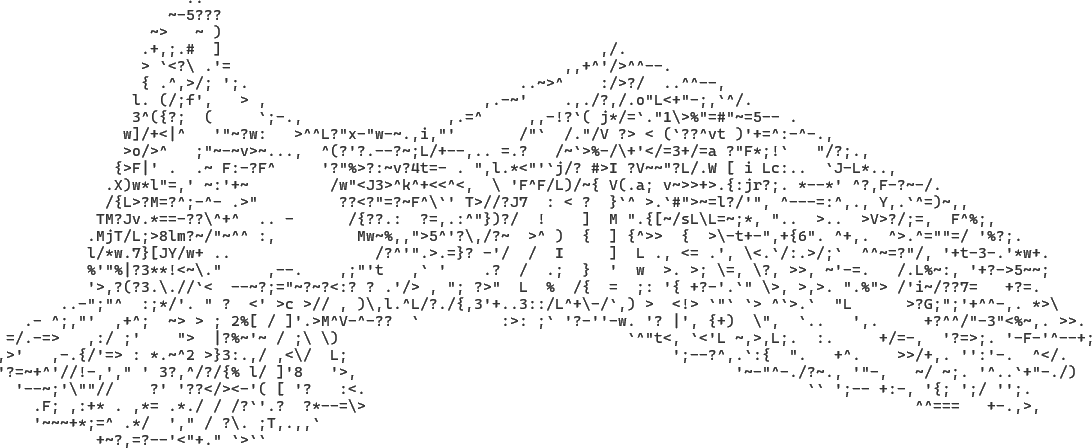}\label{fig:full_compare_bird.9}} &
        \subfloat[AAConverter]{\includegraphics[width=0.45\linewidth]{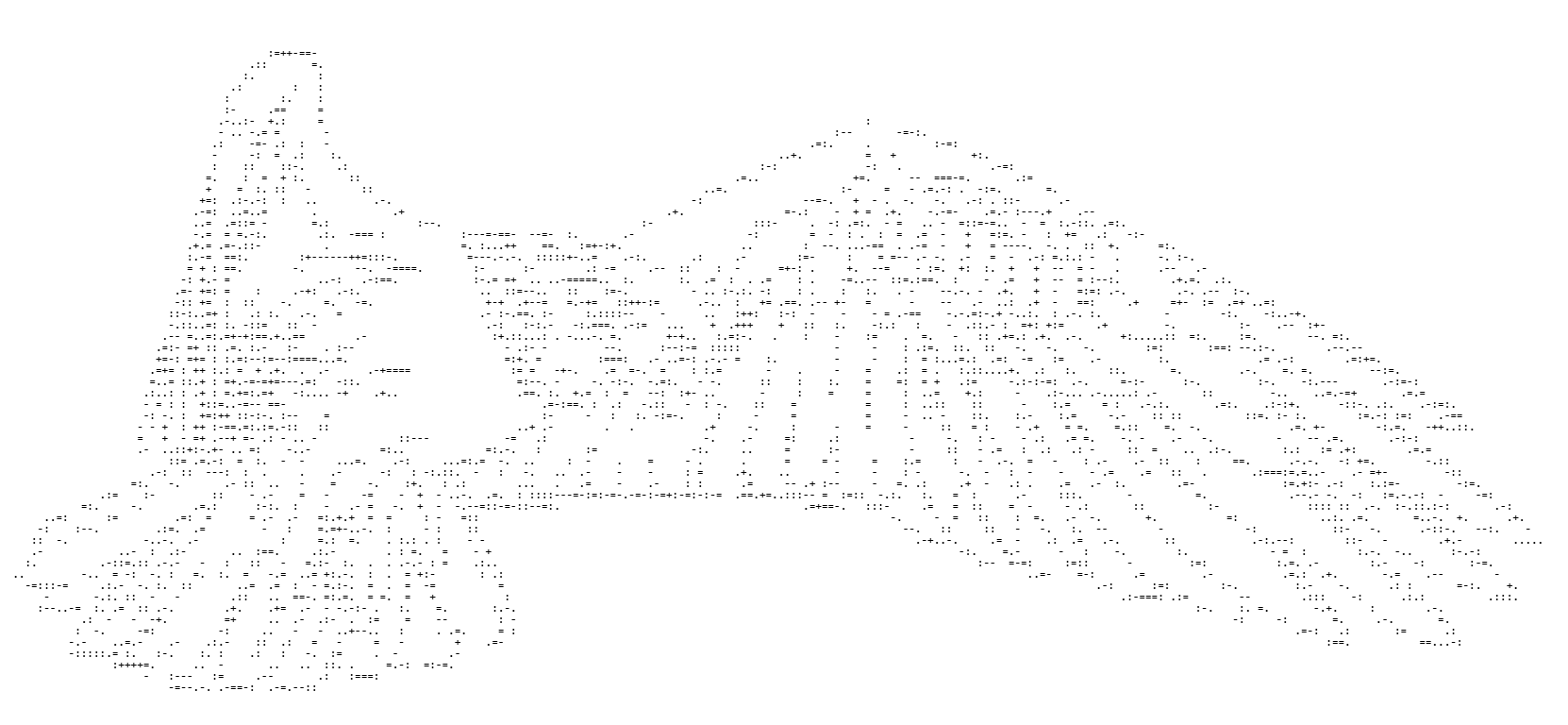}\label{fig:full_compare_bird.10}} \\
    \end{tabular}
    \caption{Detailed comparison of synthetic structure-based ASCII art for an image of a bird.}
    \Description{Detailed comparison of synthetic structure-based ASCII art for an image of a bird.}
    \label{fig:full_compare_bird}
\end{figure*}

\appendix

\end{document}